\documentclass[12pt]{JHEP3}
\usepackage{amsfonts,amssymb}
\usepackage{amsmath}

\usepackage{epsf}

\def\Y{\mathbf{Y}}

\def\T11{{T}^{1,1}}
\def\bear{\begin{eqnarray}}
\def\eear{\end{eqnarray}}

\def\dim{\mathrm{dim}}

\newcommand{\pa}{\partial}
\newcommand{\tr}{{\rm tr}}
\newcommand{\comment}[1]{}

\newcommand{\pasl}{\pa\kern-.55em /}

\newcommand{\ksl}{k\kern-.55em /}

\DeclareFixedFont{\xiiss}{OT1}{cmss}{m}{n}{12}
\DeclareFixedFont{\ixss}{OT1}{cmss}{m}{n}{9}
\DeclareFixedFont{\cmrnine}{OT1}{cmr}{m}{n}{9}
\newcommand{\field}[1]{\mathbb{#1}}

\newcommand{\BC}{{\field C}}

\newcommand{\CCs}{\hbox{\ixss C\kern-.4emI}}
\newcommand{\ZZs}{\hbox{\ixss Z\kern-.4emZ}}

\newcommand{\diag}{\hbox{diag}}

 \newcommand{\myfig}[3]{\begin{figure}[ht]
\begin{center}
\leavevmode
\epsfxsize=#2cm
\epsfbox{#1}
\end{center}
\caption{#3}
\label{fig:#1}
\end{figure}}

\title{A toy model for the AdS/CFT correspondence}
\author{David Berenstein\\ 
School of Natural Sciences, Institute for Advanced Study, Einstein Drive,
Princeton, NJ  08540, USA\\ and\\
Department of Physics, UCSB, Santa Barbara, CA 93106\\
Email: \email{dberens@physics.ucsb.edu}}
\abstract{ We study the large $N$ gauged quantum mechanics for a single Hermitian matrix in the Harmonic oscillator potential well as a toy model for the AdS/CFT correspondence. We argue that the dual geometry should be a string in two dimensions 
with a curvature of stringy size. Even though the dual geometry is not weakly curved, one can still gain knowledge of the system from a detailed study of the open-closed 
string duality. We give a mapping between the basis of states made of traces (closed strings) and the eigenvalues of the matrix (D-brane picture) in terms of Schur 
polynomials. This is interpreted as an exact open-closed duality. 
We connect this model with a decoupling limit of ${\cal N}=4 $ SYM and 
the study of giant gravitons in 
$AdS_5\times S^5$. We show that the two giant gravitons that expand along $AdS_5 $ and
$S^5$ can be interpreted in the matrix model as 
taking an eigenvalue from the Fermi sea and exciting it very much, 
or as making a hole in the Fermi sea respectively. This is similar to recent studies of the $c=1$ string. This connection gives new insight on how to perform calculations for giant gravitons.}

\preprint{hep-th/0403110}

\begin{document}

\section{Introduction}

Recently the study of the $c=1$ string theory has received a lot of 
attention, especially due to the work of McGreevy and Verlinde \cite{McV,McTV}, where a
D-brane interpretation of the dual matrix model was discussed.  Their work was partly based 
on Sen's observation that D-brane decay can be studied exactly in $\alpha'$ corrections
\cite{Sen} and the results of Fateev, Zamolodchikov and Zamolodchikov \cite{FZZ} on Liouville 
boundary field theory, where a description of the boundary states of the model was 
discussed.  Many of these results have been made more precise in the literature
\cite{KMS,TT,DKKMMS}, including a space-time interpretation in terms of type O strings for 
the solution to the non-perturbative instability of the original matrix model \cite{P}.
The $c=1$ string 
corresponds to the large $N$ limit of 
matrices on the inverted (upside down) 
harmonic oscillator. 
A review of the early literature can be found in \cite{Kleb}, and a more recent review can be found in \cite{Muk}. 
The model can be nicely described in terms of the eigenvalues of a 
Hermitian matrix $X$, which become fermions in the quantum theory \cite{BIPZ}. 
The double scaling limit that gives 
rise to the $c=1$ model is done in such a way that the 
Fermi level is close to the top of the potential, so that one can focus on the physics of 
the top of the hill.
For this model the spectrum is continuous, and the 
observables of the model can be interpreted as giving rise to an 
S-matrix. The stringy 
states correspond to small ripples on the Fermi sea which scatter from the top 
of the hill and go back to infinity, while the dual geometry is a two dimensional 
string with a linear dilaton background.
The matrix model is exactly solvable, so this matrix model
provides a 
holographic description of quantum geometry on a space-time which has an 
asymptotically flat region. The interpretation in terms of D-branes means that 
the model is gauged, so one only considers gauge invariant states, which depend only on the eigenvalues of the matrix $X$.
A lot of insight can be gained from studying the model in the phase 
space of the eigenvalues of the matrices, especially since one can give a very 
pictorial description of the model.

An equally solvable model, is the study of the large $N$ limit of the 
gauged ordinary harmonic oscillator. In general we can choose a more 
complicated potential, where the Lagrangian is given by
\begin{equation}
L= N \int dt \tr[\frac12(D_t X)^2- V(X)] 
\end{equation}
where $V$ is an arbitrary potential. For general $V(x)$ it is not possible 
to solve for the energy levels exactly, so we will concentrate on a particularly simple solvable model, where one can solve the system explicitly in more than one basis. 
Moreover, we will later show that 
the harmonic oscillator potential is special also because it 
appears as a decoupling limit of the ${\cal N}=4 $ SYM theory. This feature makes it  clear 
that this one 
particular potential originates from a bona-fide string theory in ten dimensions, and may be interpreted as a string theory in its own right.

For the quadratic potential the only tunable parameter is $N$, so 
the effective expansion in planar diagrams is the ordinary 't Hooft 
expansion \cite{tH} and the string coupling constant is $1/N^2$. 
The expansion in this case does not affect the energies of the states, but it 
does affect their overlaps.
Usually only theories in a double scaling limit are considered as string theories, but 
then,
it is not usually assumed that the $U(N)$ symmetry is gauged.
In spite of the fact that the theory seems to be free, 
one can try to give a string theory interpretation of the model.
This might turn out to be very topological in the end, as 
one does not fill the holes of  Riemann surfaces with interactions.
At the moment I do not have a good description of how to interpret this $1/N$
expansion in terms of a string worldsheet theory. In this paper this issue will not be 
explored. We will just trust that the $1/N$ expansion of 't Hooft always has such an interpretation.

A prominent feature of this model is that it has a discrete 
spectrum. In light of this fact, if this were to be interpreted as the 
holographic dual on some geometry, then the discreteness of the 
spectrum of states resembles 
the spectrum of dimensions of local operators in a conformal field theory, 
and should be viewed as giving the holographic dual of an AdS-like 
space-time in global coordinates.  Indeed, we will interpret this model as 
an example of the AdS/CFT correspondence \cite{M,W,GKP} which can be solved 
exactly. This has been proposed before as a toy model for AdS/CFT \cite{CJR}, but the proposal was not elaborated upon.  The study of free fields as a route to AdS has also been explored in
\cite{G1,G2}, although there the purpose was to write the perturbation theory in Feynman diagrams so that it resembles propagation of fields in AdS. Here instead we take the 
free model as describing the dual $AdS$ geometry. 

From the spectrum of the theory one sees that the theory does not have a Hagedorn 
growth of states. At weak coupling  and weak curvature this implies that the target space dimension is less than or equal to two. Given that we have a time variable on the boundary, holographic reasoning
tells us that we should at least include one more dimension (the radial direction on AdS), so 
that the target space of the string would have at least two dimensions. These arguments point
in opposite directions and single the target space dimension of the string theory as being equal to two.

If we assume that the description resembles the behavior of an $AdS_2$ spacetime close to the 
conformal boundary of the associated spacetime, and that the dilaton has 
some asymptotic value which is fixed by the boundary conditions, 
then to saturate the string beta functions and to obtain a critical 
string, the curvature of the spacetime needs to be of order 
$\alpha'$. This will cancel the contribution to the dilaton tadpole due to the non-critical dimension with the curvature of the embedding geometry.
Strominger has also proposed a matrix model for $AdS_2$ \cite{St} where similar features have been discussed. See also \cite{VS}.
This precludes a straightforward geometric interpretation, 
as there are no regions in the geometry 
which are weakly curved compared to the string
scale, were a semiclassical analysis would help us resolve geometry. Because of this 
issue, the dimension of the target space for the string can not be determined for certain.
 This matrix model has appeared before in the study of two dimensional black holes \cite{Ho}. There, Ho has argued that the matrix quantum mechanics described above is a limit $M\to 0$ of a two dimensional black hole and it is related to the standard $c=1$ matrix model.
 This seems difficult to achieve if we want to insist on keeping $N$ finite but large. We will leave the target space geometric interpretation of the model as an open problem.

So, even if we don't understand the target space geometry, we can anyway study the system and gain insight from other points of view. We will try to 
understand the open-closed duality in as much detail as possible.
After all, this is one way to understand the AdS/CFT correspondence when the spacetime 
geometry is highly curved \cite{KV}.  
In this sense, the gauged matrix harmonic oscillator 
should be a perfectly good toy model for the AdS/CFT.

The objective of the paper is to explain  features of the $AdS/CFT$ correspondence that 
can be realized in this matrix model, even in the absence of a string theory dual geometry. 
We find various ways to describe the spectrum of the matrix 
model exactly and relate them to each other, so that we have an open-closed duality in the 
system where all calculations can be performed. We also relate the model to ${\cal N}=4 $ SYM as a decoupled sector that describes half-BPS states of the theory. With this identification we can relate BPS states in the SYM theory
and states in the matrix model. In particular, we find that the half-BPS D-branes in SYM theory (giant gravitons) can be identified with particular configurations in the matrix model which in examples of the $c=1$ matrix model would also be called D-branes.

The plan of the paper is as follows.
In section \ref{sec:closed} we describe 
the spectrum of the matrix model in terms of ``closed string" states. This is, in terms of single trace operators, in the spirit of \cite{W}. 
Next, in section \ref{sec:eigen} we describe the spectrum in terms of the eigenvalues of $X$. 
We will call this picture the D-brane picture. In section \ref{sec:schur} we describe a new 
basis for the closed strings in terms of Schur polynomials. This description follows from
the work \cite{CJR} where all the combinatorial description is laid out in detail. Here it is shown that these Schur polynomials capture the dynamics of the eigenvalues directly. We 
give a sketch of a proof by 
comparing the wave functions of these states in a particularly simple 
regime.

In section \ref{sec:giants} we describe how this model relates to ${\cal N}=4$ SYM theory 
as a decoupled sector of ${\cal N}=4$ SYM  and 
we find applications of this new correspondence 
to the study of giant gravitons in AdS space. We find that the
two giant gravitons expanding into $AdS_5$ and $S^5$ correspond in the matrix model to taking an eigenvalue from the top of the Fermi sea of eigenvalues and exciting it by a large amount so that it is resolved from the Fermi surface, while the other giant graviton expanding into $S^5$ translates  
to making  a hole state deep in the Fermi sea of eigenvalues. This behavior
is exactly the same as the description of D-branes in the $c=1$ matrix model and goes a long way to explain why the corresponding operators in SYM theory behave as D-branes.

In section \ref{sec:other} we describe other interesting physics related to this matrix 
model, and in particular we give a matrix model description of why the correct 
non-planar perturbation parameter in the plane wave limit scales as $J^2/N$, where 
$J$ is the R-charge of a state.

 Finally, we review some of the results and conclude.

\section{Matrix description of the spectrum}\label{sec:closed}

The model we are studying is the large $N$ gauged 
harmonic oscillator. The theory can be solved by first
solving the full matrix model theory and then imposing the gauge 
invariance of the states. This is what we will do in the following.

For reasons which will become apparent later, 
we will call this picture 
of the dynamics the {\em closed string} picture. 

The system consists of a Hermitian $N\times N$ matrix $X$, (or with explicit
$U(N)$ indices 
$X^{i}_{j}$) with potential
$\frac12\tr(X^2)$, and kinetic term $\frac12\tr(D_t X)^2$, where 
$$
D_t (X) = \dot X + [A,X]
$$
and $A$ is the gauge connection and acts  as a lagrange multiplier (which is also a hermitian $N\times N$ 
matrix).
When $A=0$, the system reduces to a collection of $N^2$ free harmonic 
oscillators, and we write the Hamiltonian for these in terms of creation and 
annihilation operators
\begin{equation}
H = \frac 12(a^\dagger)^{i}_j a^j_i + \frac 12 N^2= \frac 12\tr(a^\dagger a)+\frac 12N^2
\label{eq:Hmatrix}\end{equation}
where we have left the zero point energy of the system included.

Fixing $A=0$ is a gauge choice. The only remnant of the gauge choice is that 
we have to satisfy the equations of motion of $A$, this is, 
$\delta L/\delta A= Q=0$, and $Q$ is the charge that generates gauge 
transformations. Thus we need to impose on the spectrum  of states the gauge invariance constraint $Q=0$.

The only non-trivial commutation relation of the $a$, $a^\dagger$ 
can be written as 
\begin{equation}
[(a^\dagger)^i_j,a^k_l] = \delta^i_l\delta^k_j
\end{equation}
All other commutators between $a,a^\dagger$ vanish.

The vacuum is the unique 
state satisfying $a^j_i|0>=0$ for all $i,j$. This state is invariant under 
$U(N)$ transformations by adjoint action on $X$.

An excited state of the system (ignoring the gauge constraint) is given 
by applying an arbitrary number of matrix creation operators to the 
vacuum. Each such operator increases the energy of the state by one.

Now we want to impose the gauge constraint on the system. Each creation 
operator has one upper $U(N)$ index and one lower $U(N)$ index. If we act with 
$k$ such operators on the vacuum 
we have a state which transforms as a tensor with $k$ upper $U(N)$ indices and $k$ lower 
$U(N)$ indices. To make a gauge invariant state,  we need to contract the
tensor indices of these states 
with an appropriate invariant tensor of $U(N)$. 
These invariant tensors have to be formed 
by different possible orderings of $\delta^\mu_\nu$, which contract all the 
upper indices with all the lower indices.

The collection of states obtained this way is the set of gauge invariant states 
in the large $N$ harmonic oscillator, and these are the physical states of 
the theory.

Starting with one creation operator we can follow the contraction of 
indices and write them like matrix multiplication. The states are then 
going to be given by products of expressions of the form
\begin{equation}
(a^\dagger)^{i_i}_{i_2}(a^\dagger)^{i_2}_{i_3}\dots (a^\dagger)^{i_n}_{i_1}
\sim \tr((a^\dagger)^n)
\end{equation}
These single trace states are identified with closed 
string states in the AdS/CFT correspondence \cite{W}, so we will call 
these operators the closed string states. We will label them by their 
energy $n$. The operator that creates one closed string state of energy $n$ 
is then
\begin{equation}
\beta^\dagger_n= A_n\tr((a^\dagger)^n)
\end{equation}
where $A_n$ is an appropriate normalization factor. 
For $n$ fixed, and in the 
large $N$ limit $A\sim N^{-n/2}$.

The normalization is found by studying the norm of the state as follows
\begin{equation}
|\tr((a^{\dagger})^n|0>|^2 = <0|\tr(a^n)\tr((a^\dagger)^n)|0>
\end{equation}
and this can be calculated by using Wicks theorem (free field contractions).
Explicit results for the appropriate normalizations
have been found to all orders in $1/N$ for all $n$ in \cite{KPSS}
in which they needed explicit expressions to understand the light cone string theory 
in the plane wave geometry \cite{BMN}. See also \cite{BN,C7} for related calculations. 

One can create multi string states by acting with various of these oscillators 
in succession. It is clear that $[\beta^\dagger_n,\beta^\dagger_m]=0$, so 
the spectrum of the theory resembles a Fock space of states, where there 
is one closed string oscillator per positive integer $n>0$. It is a well
known but non-trivial fact that states with different ``closed string'' 
occupation numbers are approximately orthogonal in the large $N$ limit, so 
long as we keep the energy finite when we take the $N\to \infty$ limit.
A lot of the detailed $1/N$ expansions for normalizations of the states and overlaps
can be copied verbatim 
from the study of $1/2$ BPS operators in the ${\cal N}=4$ SYM, and we will return to 
this issue later in the paper.

Given these states, we can always order the string states in descending order, so that a 
multi-string state
\begin{equation}
|n_1,n_2,\dots,n_k> = \beta^\dagger_{n_1}\dots\beta^\dagger_{n_k}|0>
\end{equation}
satisfies $n_1\geq n_2\geq n_3\dots\geq n_k$. The total energy of the 
state above the ground state energy $\frac 12 N^2$ is 
$\sum n_k = n$. The number of states with energy $n$ is given by the partitions 
of $n$ into positive integers for large $N$. At finite $N$ one needs to remember 
that traces of different length are not algebraically independent, indeed 
$tr((a^\dagger)^{N+1})$ can be written as a polynomial of traces of lower length.

In the large $N$ limit, the spectrum constructed above coincides with the spectrum of 
a chiral boson in $1+1$ dimensions. This point of view agrees with our description of 
the target space  geometry in the introduction. The model suggests that the dual target 
space geometry has one field theory degree of freedom (this would be the ``tachyon", as dilatonic  gravity 
has no propagating degrees of freedom in two dimensions).

At finite $N$, the spectrum gets cut (this is called the stringy exclusion principle, which is non-perturbative in $N$), and the spectrum is determined by partitions of $n$ into integers smaller or equal to $N$. To each configuration of traces we can associate a Young tableaux. We first order the integers in the multi-trace state
so that they are decreasing. This is, we label the state
\begin{equation}
\beta^\dagger_{n_1}\dots \beta^\dagger_{n_k}|0>
\end{equation}
with $N\geq n_1\geq n_2\dots\geq n_k$ by a Young tableaux where the first column has $n_1$ boxes, the second column has $n_2$ boxes, etc. And the maximum length of each column is $N$.

\section{The eigenvalue basis}\label{sec:eigen}

Now, we will look at a second gauge choice, where we choose the matrix $X$ to be 
diagonal. In the $c=1$ matrix model the
eigenvalues represent D-branes. Here, we will use this same interpretation, so a description in terms of eigenvalues will be the description in terms of D-branes. This will be an 
{\em open string} description of the system.

 Let us label the eigenvalues of $X$ as $\lambda_i$.
 Then, when we write wave functions for the Schr\"odinger equation, they will be functions of $\lambda_i$. There is a discrete subgroup of $U(N)$ which leaves the matrix $X$ diagonal. This is the permutation group of the eigenvalues, so the wave functions have to be invariant under this 
symmetry, and this means that we get totally symmetric wave functions on the 
eigenvalues.

Classically, the Lagrangian for the eigenvalue basis becomes 
\begin{equation}
L = \sum \frac 12\dot\lambda^2_i-\frac 12\lambda_i^2
\end{equation}
So the classical motion of the eigenvalues is that of a harmonic oscillator. 
However, quantum mechanically there is a change of measure from the matrix basis to the eigenvalue basis. This change of measure is the volume of the gauge orbit of the matrix $X$, and it is equal to the square of the Van der Monde determinant of the $\lambda_i$, namely
\begin{equation}
\mu = \Delta(\lambda)^2 = \prod_{i\neq j}(\lambda_i-\lambda_j)
\end{equation} 
So that the Hamitonian in the quantum theory will be given by 
\begin{equation}
H \psi = \frac12 \sum -\mu^{-1} \partial_{\lambda_i} (\mu \partial_{\lambda_i} \psi) +\lambda^2_i \psi
\end{equation}
with $\psi$ the wave function of the eigenvalues.

The measure can be absorbed in the wave functions for the $\lambda_i$, by 
attaching a factor of the Van der Monde to the wave function. We define $\psi(\lambda) = \Delta^{-1}(\lambda)\tilde\psi(\lambda)$, where $\tilde\psi(\lambda)$ is the new wave function in the $X$ variables expressed in terms of the eigenvalues of $X$ (these are the $\lambda_i$), and the measure for $\tilde\psi$ is is just $\prod d\lambda_i$. This can be done for any one matrix model quantum mechanics \cite{BIPZ} with a single trace potential.  This is a similarity transformation on the space of wave functions, so it affects the form of the Hamilltonian. The new Hamiltonian is 
\begin{equation}
\tilde H = \frac 12\sum_i -\partial_{\lambda_i}^2+\lambda_i^2
\end{equation}
so it becomes a Hamiltonian for $N$ free particles in the harmonic oscillator potential well.
After this is done the wave functions are completely antisymmetric in the $\lambda_i$: the eigenvalues become fermions due to the Van Der Monde determinant.
The system is reduced to $N$ free fermions in a given potential, which for us is just $V(x) = x^2/2$. For our setup, an orthogonal  basis for 
the N-particle wave functions is given by Slater determinants of 
one particle wave functions for the Harmonic oscillator (these are in turn given by Hermite polynomials times a Gaussian factor $H^{n_k}(\lambda)\exp{-\lambda^2/2}$). This basis for the wave functions is given explicitly by
\begin{equation}
\psi(n_1, \dots ,n_N) \sim \det\begin{pmatrix} H^{n_1}(\lambda_1)&H^{n_1}(\lambda_2)&\dots&H^{n_1}(\lambda_N)\\
H^{n_2}(\lambda_1)&H^{n_2}(\lambda_2)&\dots&H^{n_2}(\lambda_N)\\
\vdots&\vdots&\ddots&\vdots\\
H^{n_N}(\lambda_1)&H^{n_N}(\lambda_2)&\dots&H^{n_N}(\lambda_N)
\end{pmatrix}\exp(-\sum \lambda^2/2)
\end{equation}
In particular, the Fermi statistics imply that all of the $n_k$ are different, and that 
we can order the $n_i$ so that $n_1>n_2>n_3>\dots>n_N\geq 0$.
The energy of a state is then $\sum_i (n_i+1/2)$.
The ground state of the system is such that the $n_i$ are minimal. 
This is, $n_{N-k}=k$. From here it follows that the ground state energy of the system 
is
\begin{equation}
\sum_{k=0}^{N-1} \frac 12(2k+1)=\frac {N^2}2
\end{equation}
which coincides exactly with the c-number term in equation \ref{eq:Hmatrix}, where the 
Hamiltonian is written in normal ordered form. We can also write the spectrum as the list of 
non-increasing integers given by 
$n'_1=n_1-N\geq n_2'= n_2+1-N\geq \dots\geq n_k'=n_k-(N-k) \geq n_N'=n_N\geq 0$. This coincides with the description of the spectrum given in terms of `closed strings' in the previous section. The difference, however, is that in the eigenvalue basis, for different values of the list of integers $n_\alpha$ we get orthogonal states.  Because of the description in terms of Fermions,  the ground state describes a Fermi sea of eigenvalues, where the level of the sea is determined by $N$. 
We can look at the spectrum of excitations as given by exciting  the Fermi surface of the Fermi sea. The highest fermion of the Fermi surface has it's energy raised by $n_1'$ units, the next to highest fermion has it's energy raised by $n_2'$ units and so on. For finite energy excitations (in the large N limit) only the topmost eigenvalues get excited beyond their ground states. One can also 
look at states whose energy scales with $N$ in some way in the large $N$ limit. This is not 
anymore the usual large 
$N$ limit of 't Hooft\cite{tH}. \footnote{These types of states will become important later. The energy of the states we will look at will scale proportionately to $N\sim g_s^{-1}= 1/((1/N))$, so they have the potential to be interpreted as D-branes, because their tension will be proportional to the inverse of the 't Hooft string coupling constant $g_s^2=N^{-2}$.}

Again, one can describe a state in the spectrum by drawing a Young tableaux. 
The tableaux is written so that the first row has $n'_1$ boxes, the second row has $n_2'$ boxes, etc. The tableaux has only $N$ rows, as there are only $N$ different 
eigenvalues.

The following two pictures fig. \ref{fig:harmosc.eps} and fig. \ref{fig:phase.eps} illustrate how one fills the Harmonic oscillator potential well with
fermions, and the description of the Fermi surface in the phase space of a single eigenvalue. The string states found in the previous section will be general small 
perturbations of the Fermi surface of the eigenvalue distribution in the the phase space of a single eigenvalue. 
The description in terms of closed strings describe collective excitations of the fermions, since the basis of states is different than the one found here, so they describe mixed states 
of fixed energy where we can not determine the energy of a single eigenvalue exactly.
 These collective excitations are interpreted as changes in the shape of an incompressible 
Fermi liquid droplet in the eigenvalue phase space. The incompressibility arises because each eigenvalue state occupies one quantum of area, and Fermi statistics forces the eigenvalues to be centered around different locations.
A single closed string state with energy $k$ is interpreted as 
a single quantum of a 
wave on the edge of the droplet with wave number $k$.

\myfig{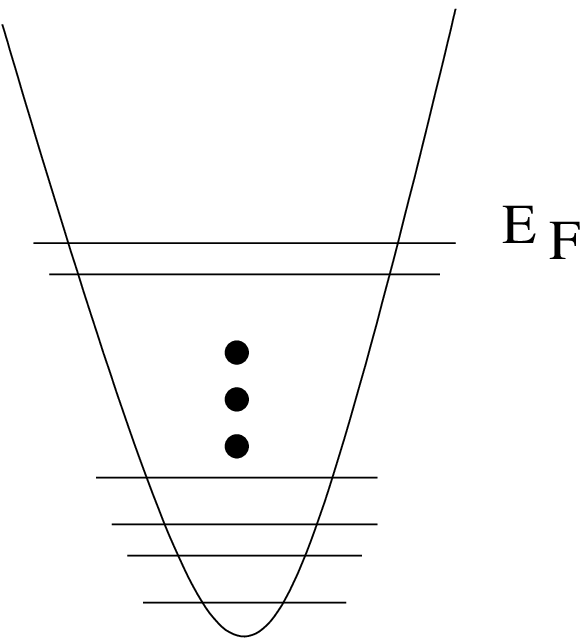}{5}{Filling the potential well}

\myfig{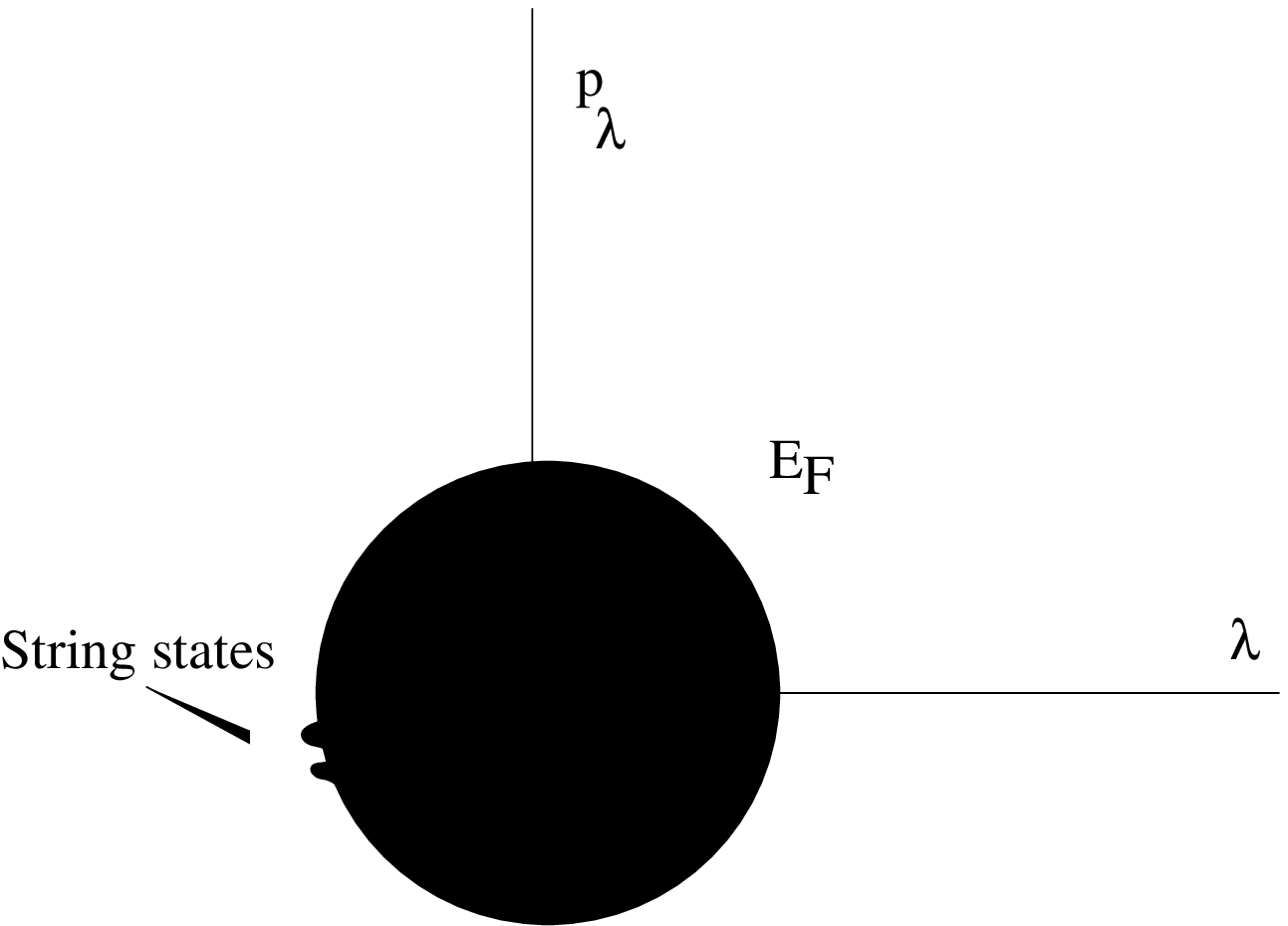}{7}{String states as small perturbations of the Fermi surface}

\section{Schur polynomial basis}\label{sec:schur}

We will now use a third description of the spectrum of the theory. This again proceeds
through choosing $A=0$, but we will write the multi-string states in a different basis.

The main idea is to use the construction of gauge invariant states  proposed in \cite{CJR}
based on Schur polynomials. The basis construction 
proceeds by writing an auxiliary space $V$ which transforms in the fundamental of 
$U(N)$. We can then think of a hermitian matrix $X$ as a linear map  
$X:V\to V$. The character of $X$ in $V$ is exactly $\tr(X)$, and this is invariant under general changes of basis of $V$, which are done by the complexification of $U(N)$, namely $GL(N,\BC)$. Now, let us consider 
the tensor product space $\Omega_n=V^{\otimes n}$, and define an action of $X$ on
$\Omega_n$ as follows
\begin{equation}
X(v_1\otimes\dots\otimes v_n) =  (X v_1)\otimes (X v_2)\otimes\dots\otimes(X v_n)
\end{equation}
If $X$ is invertible, we can think of $X$ as an element of $GL(N,\BC)$, and this is the group action of $X$ on the tensor product. If we decompose $\Omega_n$ into irreducible representations of $GL(N,\BC)$, then $X$ acts diagonally under this decomposition.
The action of $X$ commutes with the permutations of the vectors $v_i$, so it is possible
to digonalize $X$ and the permutation group simultaneously. Thus we decompose 
$\Omega_n$ in terms of representations of the permutation group of $n$ elements. This sets up a correspondence between representations of the symmetric group of $n$ elements and irreducible representations of $GL(N,\BC)$, which is described exactly by Young tableaux with $n$ boxes. Thus we can associate to each Young tableaux a group representation of $GL(N,\BC)$ which sits inside $\Omega_n$, 
and an associated action of $X$ on that same representation which is induced from 
projection of the action of $X$ on $\Omega_n$. Said more simply,  if we restrict to tensors in $\Omega_n$ of a specific 
symmetry type, then the action of $X$ on these tensors induced from the action of $\Omega_n$ preserves the symmetry type.

Symmetry types of tensors are in one to one correspondence with 
representations $R$ of $SU(N)$. In each of these representations 
we can find a gauge invariant observable which is the character of $X$ in the associated representation  $\tr_R(X)$. This is very similar to the characterization of observables in two dimensional QCD in terms of Wilson loops around non-contractible cycles, taking all possible representations of the group into account\cite{G}.
See also \cite{MP}.
 With proper normalization, these are called Schur polynomials. 
We can extend this action to matrix valued operators acting on some Hilbert space, so we can use the following basis 
\begin{equation}
\tr_R(a^\dagger)|0>
\end{equation}
as a collection of states of the large $N$ harmonic oscillator. At energy $n$ over the ground state, there are as many partitions of $n$ with less than or equal to $N$ rows 
as there are Young Tableaux representing irreducible representations of $SU(N)$. Moreover, as discussed in \cite{CJR}, these states are actually orthogonal, so one can build this way an
orthonormal basis  of states which capture all of the states of the gauged harmonic oscillator with $N\times N$ matrices.

Now, we want to ask what is the relation between the three basis of states we have discussed:
the {\em closed string basis}, the eigenvalue basis (we called this an {\em open string} description as it related to D-branes), 
and this new basis which we will call the Schur polynomial basis.

Going from the string basis to the Schur polynomial basis is straightforward, as the
projections to the different irreducible components of $\Omega_n$ are done by taking appropriate symmetrizations over rows of the Young tableaux, and antisymmetrizing 
over the columns.
For example, we can take the antisymmetric and symmetric representations 
(with two boxes) and we find that
\begin{equation}
\tr_A(X) = \frac12(\tr(X)^2-\tr(X^2))\quad\tr_S(X) =\frac12( \tr(X)^2+\tr(X^2))
\end{equation}
The states $\tr_A(a^\dagger)$ and $\tr_S(a^\dagger)$ 
mix maximally the different number of traces, so these are always 
interpreted as multi-closed string states.

The surprise is that the Schur polynomial basis seems to coincide exactly with the eigenvalue basis.
This equivalence of basis 
was hinted in \cite{JvT,CJR}, and was also found in \cite{Poly} in the study of the 
Calogero model. 
A sketch of the proof goes as follows.

Let us consider the wave functions in the eigenvalue basis. As we said previously, these are
determined by Slater determinants of Hermite polynomials times a Gaussian factor (which is common for all wave functions). Let us strip the Gaussian part of the wave function, so we are left with polynomials of the eigenvalues only. Take the limit $\lambda_1>>\lambda_2>>\dots>>\lambda_N>>1$. In this limit the wave function is dominated by the leading term (up to normalization factors)
\begin{equation}
\psi(\lambda_1,\dots,\lambda_N)\sim \lambda_1^{n_1}\dots\lambda_N^{n_N}\label{eq:asympslater}
\end{equation}
with $n_1>n_2\dots >n_N$.

Now let us consider the operators $(a^\dagger)^{\otimes n} |0>\sim 
(X+\partial_X)^{\otimes n}|0>$. The leading term in $X$ for large $X$ will be given by 
letting $\partial_X$ act on the Gaussian factor, so that we can approximate $(a^\dagger)^{\otimes n}$ by $X^{\otimes n}$ up to numerical factors.

Let us now look at the matrix $X$. We can diagonalize it and evaluate 
$\tr_R(X)$ explicitly. 
We do this by choosing $X$ to be diagonal with eigenvalues chosen in 
decreasing order, in the same asymptotic regime that we chose in the eigenvalue basis. 
Choose the Cartan of $GL(N,C)$ so that $X$ is in the Cartan.
The highest weight state $|\alpha_R>$ 
will have weights $n^R_1 e_1+n^R_2e_2+\dots$,  where the $e_i$ are
the positive roots of $SU(N)$ and the $n^R_i$ are the lengths of the rows of the Young tableaux. The character of $X$ will sum over the elements 
of the weight lattice that  belong to the representation $R$.
The leading term is $X$ acting on the highest weight state, so that
\begin{equation}
\tr_R(X) \sim \lambda_1^{n_R^1}\dots \lambda_N^{n_R^N}
\end{equation}
we also need to remember that the ground state wave function in the eigenvalue
basis has an extra leading term from the Van Der Monde determinant, $\lambda_1^{N-1}\dots \lambda_{N-1}$. Multiplying both of these and stripping the Gaussian term
we find the asymptotic behavior
\begin{equation}
\tr_R(X)|0> \sim \lambda_1^{n_R^1+N}\dots \lambda_N^{n_R^N}
\end{equation}
with $n'_1= n_R^1+N>n'_2= n_R^2+(N-1)>\dots>n'_N= n_R^N$.

We need to compare the asymptotic behavior we just found 
with the one coming from the Slater determinant wave functions in the eigenvalue basis from equation \ref{eq:asympslater}.
We notice that the two states which are associated with the same Young tableaux, have the same asymptotic behavior.

Now, we order the states according to how fast they grow in the asymptotic regime we are studying. A monomial of the form $\lambda_1^{s_1}\dots\lambda_N^{s_N}$
will have higher ordering than $\lambda_1^{t_1}\dots\lambda_N^{t_N}$ if 
it is of higher degree: $\sum s_i \geq \sum t_i$. 
In the case of equality (which will be the case for the highest degree of the wave function),
 we also require
that the smallest integer $k$ for which $s_k\neq t_k$ is such that 
$s_k > t_k$. This ordering of the monomials (from the smallest to the highest) produces a filtration 
 of the Hilbert space of 
energy eigenvalues of the same energy, which is tied to the asymptotic growth of the wave function. A filtration is a collection of subvector spaces
$V_1\subset V_2 \subset V_3\subset\dots \subset H$. Here the $V_i$ are
ordered by the asymptotic behavior of the most divergent term in the wave function.
A wave function $\psi\in V_i$ is such that  $\psi$
diverges at most as fast as the associated monomial associated to $V_i$. The vector space quotients satisfy $\dim (V_{i+1}/V_i ) = 1$. This ordering of monomials translates to 
an ordering of the Young tableaux, so that $V_k$ is a vector space of dimension $k$ 
generated by the $k$ smallest young tableaux with a fixed number of boxes.

The fact that the two basis of states, associated to Schur polynomials and Slater determinants
(lets us call them $E_1, \dots E_k$ and $
E_1', \dots E_k'$) have the same asymptotic behavior 
means that \begin{equation}
V_i= \hbox{span} (E_1,\dots E_i) = \hbox{span} (E_1',\dots E_i')
\end{equation}
and in particular, we get that since $\dim(V_1)=1$, then $E_1= E_1'$ (up to normalization).  
Also, the $E_\alpha$ and $E_\alpha'$ are an orthogonal basis for each $V_i$, so that 
$V_i= V_{i-1}\oplus \hbox{span} (E_i)= V_{i-1}\oplus \hbox{span} (E'_i)$
where the decomposition is in terms of orthogonal complements. It follows that 
$\hbox{span} (E_i)= \hbox{span} (E'_i)$, and since this is a one-dimensional 
vector space, we have that $E_i= E_i'$ (up to normalization).
This shows that the two basis of states are equivalent.

In essence, the description in terms of Schur polynomials and the eigenvalue basis coincide.
With this information we reduce the problem of three basis to two, and the 
relation between them is very explicit. From our point of view, this is an exact 
open-closed string duality and this result should be viewed as establishing
a very natural setup for the AdS/CFT correspondence.

\section{Relations to ${\cal N}=4 $ SYM }\label{sec:giants}

We now want to see that the matrix model we have been studying appears as a 
decoupling limit of ${\cal  N}=4 $ SYM theory. This will provide further 
evidence that this model has a string theory interpretation.

Consider a time slicing of
 $AdS_5\times S^5$, similar to the ideas in \cite{BMN}, so that the Hamiltonian is
 given by 
 \begin{equation}
 H_\epsilon = \frac{(\Delta-J)+\epsilon\Delta}{\epsilon}
 \end{equation}
where $\Delta$ is the dilatation operator, and $J$ is one of the R-charges of ${\cal N}=4$ 
SYM.

Take now $\lim_{\epsilon\to 0} H_{\epsilon}$ and we find that for any state 
where $\Delta-J>0$ \footnote{
The BPS inequality is $\Delta-J\geq 0$} 
the Hamiltonian gives a very large energy, so these states can be decoupled from
the low energy theory. The only states which remain are the half BPS states 
of $AdS_5\times S^5$. It can be easily seen that in the free field theory limit of ${\cal N}=4 $ SYM theory, all other states which are not half BPS with respect to the R-charge $J$ have $\Delta-J\geq 1$, so that they carry very large energy with respect to $H$. Even in the presence of interactions we expect that these states will not suddenly become very light, so that a decoupled sector remains, as the low-lying states are protected by supersymmetry and are not lifted from having zero energy. 

We will now argue that the description of this limit gives exactly a one matrix quantum mechanical system (the 
matrix will be complex, but it is characterized by the same number of states as we have been discussing in the rest of the paper.)

The first thing we need to establish is a way of comparing the hermitian matrix 
model results with BPS operators in ${\cal N}=4$ in SYM theory. The correspondence 
proceeds as follows. Take SYM on $S^3$ and choose the gauge in the classical vacuum so 
that $A=0$. Now, we decompose all fields in their spherical harmonics. For the 
complex scalars $(\phi^i)^j_k$ one has a singlet under the $SO(4)$ symmetry group, which is a constant mode on the $S^3$, this spherical harmonic is $\phi^i_0$. 
 This corresponds to the local operator $\phi^i_j(0)$ in the 
operator state-correspondence. The other spherical harmonics are
states which transform non-trivially under the $SO(4)$ of rotations,
 and these are given in the local operator language by covariant derivatives of the field $(\phi^i_j)_\alpha \sim 
D_\alpha (\phi^i_j)$, where we think of $D_\alpha$ as a derivative operator of order 
$|\alpha|$ depending on the appropriate representation of $SO(3)$ which is completely symmetric in the derivatives (there is after all a non-trivial commutation relation between covariant derivatives which leads to ambiguities of ``normal ordering" of the derivatives.)
This definition has the correct free field theory limit.

In this free field limit, the quantization produces one harmonic oscillator per mode on the 
sphere $S^3$. The complex field for a given spherical harmonic will be a linear combination of creation operators for the field $\phi^i$ and annihilation operators for the
field $\bar\phi^i$. The dictionary then states that \begin{equation}\phi^i_\alpha\sim
Y_\alpha (a^i)_\alpha^\dagger+ Y_\alpha ( \bar a^i)_\alpha\end{equation}
 We will consider only half BPS operators where the $SO(6)$ R-symmetry is broken down to $SO(4)$. This is, we will be interested in half BPS states which are 
 highest weights of $SO(6)$. These are operators that depend on only one complex 
 scalar of the ${\cal N}=4$ multiplet, let us call it $\phi$.

When we take the operator $\tr(\phi^n)(0)$ or $\tr_R(\phi)(0)$, which is a half-BPS state,  
we are instructed to take 
the state $\tr((\phi_0)^n)|0>$ 	or $\tr_R(\phi_0)|0>$ on $S^3$, where we 
are restricted  to the S-wave on the $S^3$. 
Notice that because these 
states are made of spherically invariant oscillators, we do not need to worry about the gauge fields which are non-spherically symmetric. The only spherically symmetric gauge field is 
the S-wave of $A_0$. This field is non-dynamical,  but it 
also survives the limit and is required to implement the gauge invariance constraint on 
the allowed sates coming from the spectrum of ${\cal N}=4$ SYM theory.  

From the ${\cal N}=4 $SYM theory, when we choose the different time slicing,  
we just keep the 
creation operators for quanta of $\phi$ and not $\bar\phi$, so we get the same 
Hamiltonian as we studied. As argued above, all other states become very 
massive, so we can ignore them.
Even in the interacting field theory case there are no higher 
polynomial interactions of the field $\phi$ that do not involve $\bar \phi$ or other fields:
 all of these are
set to zero because we are on the ground state for the $\bar \phi$ oscillators. 
The description of the half BPS states then requires us to choose the matrix model with harmonic oscillator potential.

The reader might complain that the matrix $\phi$ is 
complex and the $U(N)$ gauging is not sufficient to diagonalize it. However, in 
supersymmetric field theories the gauge group is usually complexified. Also, we are only 
keeping half of the degrees of freedom of the complex matrix pair $\phi,\bar\phi$ so in 
the end we have the same number of dynamical degrees of freedom as the model 
studied in this paper.
At least formally, this produces a decoupled sector of the AdS/CFT 
correspondence which is consistent, as all other degrees of freedom are integrated out 
because they cost too much energy.

 A description in terms of a complex matrix model is also natural if one views 
 the eigenvalue phase space as 
the description of the lowest Landau level of a 2D fermion in a magnetic field.
In this case the coordinates $X,Y$ describing the degeneracy of the landau levels do not commute, and can be associated to the phase space of a single coordinate $X$.  This is how we can relate the model to the quantum Hall effect. Indeed, recently it has been argued in 
\cite{BKR} that how one chooses to order the states in the 2D system is
analogous to choosing a time coordinate in general relativity. Taking 
$Z=X+iY$ and $\bar Z = X-iY$ as the phase space coordinates produces a complex matrix 
model with a single complex coordinate $Z$, which is clearly equivalent to the matrix model for $X$ alone. 

Since the annihilation operators $\bar a$ act trivially on the vacuum, and since for the complex scalar field there are no self-contractions, the gauge invariant states that we can build with finite energy are just of the form that we described in sections \ref{sec:closed} and \ref{sec:schur}, namely either traces or Schur polynomials of a unique matrix creation operator acting on the vacuum.
In this way, we should be able to interpret the 
states created by Schur polynomials in terms of the eigenvalues of the complex matrix 
$\phi$.

Of course, the study of half-BPS objects is interesting only if there are nice 
configurations which can be interpreted geometrically on 
$AdS_5\times S^5$ and we want to understand the AdS/CFT dictionary. 
Such objects exist, and they have an interpretation as dynamically stable D-brane solutions: giant gravitons.
Giant gravitons \cite{McST} are 
D-brane solutions found in $AdS_5\times S^5$  which  wrap an 
$S^3$ and spin on $S^5$ and which preserve half of the supersymmetries. Hence they have the same quantum numbers as gravitons. These were used to explain the stringy exclusion 
principle: in this case, the fact that there is an upper bound on the angular momentum of a single string state, namely $\tr(\phi^N)$. Later it was found that there are other D-brane
objects, giant gravitons which expand on $AdS$ and which are also spinning on the $S^5$ which have the same quantum numbers \cite{GMT, HHI} where there is no upper bound on the 
angular momentum that they carry. In the dual ${\cal N}=4$ theory, they should be represented by some local operator. 
In the papers \cite{BBNS} and \cite{CJR} it was proposed that there are two types of 
operators which correspond to having giant gravitons in the dual spacetime. 
The two operators which were conjectured to be dual to these two D-brane configurations
 are given by $\tr(\phi)_R$ for two very simple
Young tableaux: the ones with one column (totally antisymmetric representations of 
$GL(N,\BC)$ \cite{BBNS}) or the ones  with one row (totally symmetric representations of $GL(N,\BC)$\cite{CJR}). Their evidence for these operators corresponding to D-branes 
was that operators made of traces mix too much to be useful, so operators with better orthogonality properties should do the trick. 

Now, let us turn to the matrix model and see what these operators do in the eigenvalue basis. An operator 
which is totally symmetric with a number of boxes $m$ of order $N$ is identified 
with the state $\tr_R(a^\dagger)$, where $R$ is the associated Young tableaux to the totally symmetric representation: one row of boxes. We previously saw that this description in terms of Young tableaux corresponds exactly to the eigenvalues basis. Thus, a totally symmetric representation corresponds to taking 
the topmost eigenvalue of the Fermi sea and giving it an additional energy  $m$ which is large. This configuration is thus an eigenvalue very far from the Fermi sea. This is the picture 
of D-branes found in the $c=1$ matrix model \cite{McV,McTV}. This is a very natural description also in light of the ideas for Matrix theory in \cite{BFSS}. The original description as an ``operator" state makes it look like a giant graviton can only be understood in a very ``quantum mechanical" way in terms of the theory on the boundary. This shows that there 
is a way to think about this D-branes in a more traditional sense.
 The second type of operator, a giant graviton expanding on $S^5$ corresponds to a Young tableaux which is a 
column with $m$ boxes, and $m\leq N$ but of order $N$. This corresponds to taking 
the $m$ topmost eigenvalues and giving to each one quantum of energy. This creates 
a hole deep in the Fermi sea of eigenvalues of the matrix model, which corresponds to another type of D-brane state
which was also described in the $c=1$ matrix model and the boundary Liouville field theory in \cite{DKKMMS,GIR}. So from the $c=1$ matrix model perspective it is natural that these two objects behave like D-branes.  In principle we could have started in the opposite direction and discovered the giant graviton operators.

One should contrast this intuition with the much more cumbersome combinatorial techniques that were used in \cite{AABF,B} to show that these states have a well defined $1/N$ 
expansion and can accommodate a spectrum of open strings. However, some of these results went beyond the study of half BPS operators alone.
At least if we restrict 
to half BPS objects, we can identify states easily in the Young diagrams that correspond 
to open string excitations and closed string excitations very explicitly. See the figures \ref{fig:ocYoung.eps} and \ref{fig:phoc.eps}.

\myfig{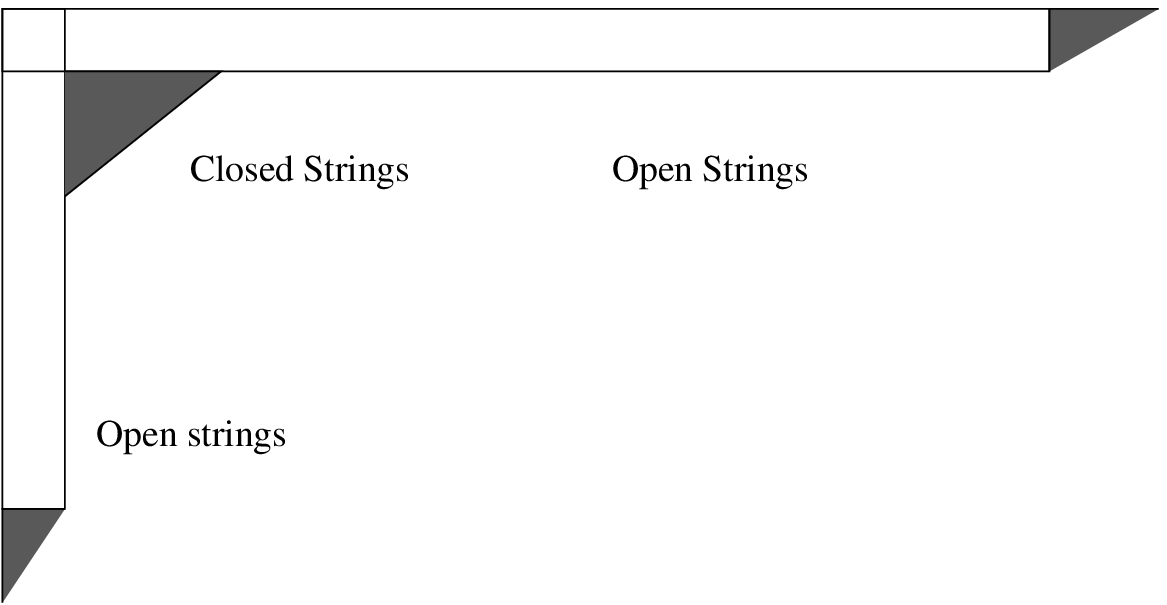}{10}{Young diagram identifications of open and closed strings}

\myfig{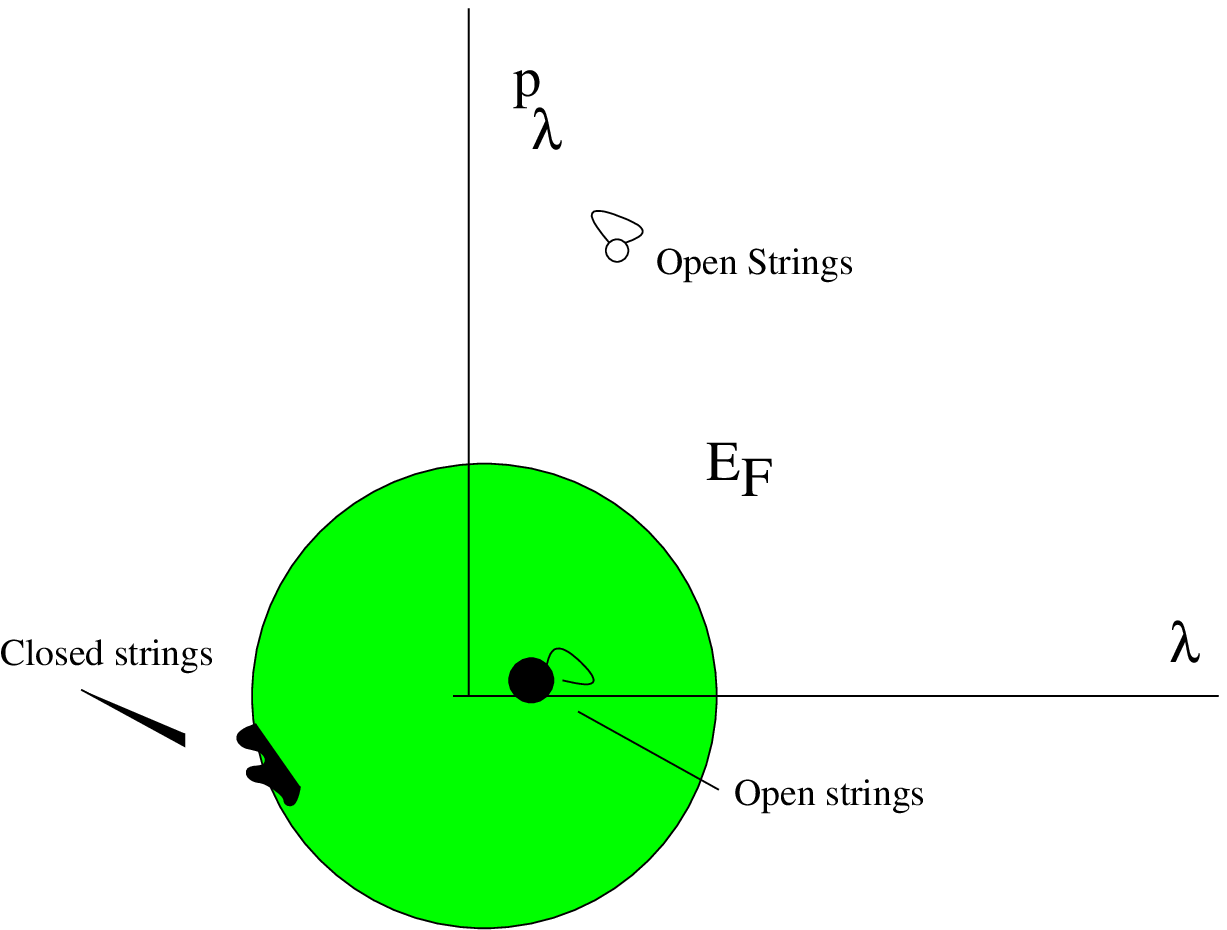}{8}{The interpretation of the Young diagram 
strings in the eigenvalue phase space}

The physical description is as follows: to change the position of the lone eigenvalue or the hole we add or subtract a finite number of boxes from the column or row. These operations are interpreted as exciting the open strings from the D-brane to itself.
To add small excitations to the Fermi sea we act on the topmost eigenvalues of the Fermi sea 
and add some boxes.

Here, from the matrix model point of view the stringy exclusion principle has a different interpretation: the Fermi sea is not 
infinitely deep. Thus  holes in the Fermi sea have a bound on their energy. These are the giant gravitons that expand into $S^5$. 

Also, the Young tableaux lets us visualize the gauge symmetry enhancement when two D-branes come together. This is shown schematically in figure \ref{fig:ocYoung.eps} where the
open string excitations are drawn in triangular form. This is the constraint on Young 
tableaux from the fact that the rows have non-increasing length. If we start with $k$ 
equal rows and add boxes, we get that the excitations coincide with the set of gauge 
invariant operators for a $U(k)$ gaussian matrix quantum mechanics. The same is true for 
the holes. If on the other hand we have two D-branes with very different energies, we 
can add excitations to each of them independently of the other one. This can be interpreted
as the Higgs mechanism, where the $U(2)$ gauged symmetry is broken to $U(1)^2$ and the D-branes become very separated from each other. Then we only have fluctuations of the positions of each D-brane independently of the other one.

The new point of view on giant gravitons also helps in defining a new way to do calculations 
with giant gravitons: a semiclassical calculation. This is in line with the observation of \cite{GKP2} that any time that a large quantum number appears, the result can be understood
in terms of semiclassical physics. In this case, it is the semiclassical physics of the
single  eigenvalue which generates the giant graviton. Based on the AdS dual description of 
the giants it was conjectured that this was the case \cite{HHI} because the RR flux inside the
giant was reduced by one.

The simplest case is that of a single giant graviton,  when we excite one eigenvalue very much.
We need to keep track of both the field $\phi_0= X+iY$ and $\bar\phi_0=X-iY$ 
to make this semiclassical calculation explicit. Indeed, in terms of $X,Y$ the effective 
Hamiltonian for the free field theory limit is
\begin{equation}
\frac12 (\tr(p_X^2+X^2+p_Y^2+Y^2))
\end{equation}
while the R-charge angular momentum is $J=\tr(p_x Y - X p_y)$.
The general classical 
solution for one excited eigenvalue of  $X$,$Y$ is \begin{eqnarray}
X&=& \diag(\lambda,0,0,\dots,0)\sin(t+\varphi)\\Y&=&\diag(\lambda',0,
0,\dots)\sin(t+\chi) \\
p_X&=&\diag(\lambda,0,0,\dots,0)\cos(t+\varphi)\\
p_Y&=&\diag(\lambda',0,
0,\dots)\cos(t+\chi)
 \end{eqnarray}
 while the BPS constraint is  $J=H$. The classical value of $H$ is given by 
 \begin{equation}
 H=\frac 12(\lambda^2+\lambda'^2)
 \end{equation}
and 
\begin{equation}
J = (\lambda\lambda')\sin(\chi-\varphi) 
\end{equation}
The BPS condition becomes $\varphi= \chi-\pi/2$ and 
$\lambda=\lambda'$. This means that $\phi$ only has a positive component 
frequency: $\phi = \diag(\lambda,0\dots 0) \exp(i t+\varphi)$.

In principle, one should be able to use this solution in the interacting theory to 
obtain some information on the spectrum of open strings.
The D-branes realized by holes should also have a semiclassical description, but 
we need to treat the hole very differently from the eigenvalue above.
In principle, we should aim to match results available in the literature
\cite{BHLN,B}, once a framework for doing the semiclassical calculation is found.

\section{Other interesting features}\label{sec:other}

The picture of eigenvalues in the Fermi surface also has interesting consequences for 
the string perturbation expansion. Indeed, we can ask what is the physical reason 
that in the BMN 
limit of ${\cal N}=4$ SYM \cite{BMN}, the 
non-planar perturbation expansion for overlap terms in terms of traces behaves like an expansion
in $J^4/N^2$ \cite{KPSS,BN,C7} with that particular power of $J$. The $1/N^2$ 
dependence is exactly the one expected from 't Hooft.  

From the eigenvalue picture, the wave functions of eigenvalues near the top of the Fermi 
sea behave roughly as $\lambda^n \exp(-\lambda^2/2)$, and they have a maximum for $\lambda\sim\sqrt{n}$. Expanding around the maximum we find that the value of the wave function decays for large $\lambda$ as $\exp{(-\frac12(\lambda-\sqrt n)^2)}\psi_n(\lambda_0)$, so that the thickness of the wave function for each eigenvalue is more or less uniform, of size $1$. The string states with momentum $J$ 
will create perturbations of the Fermi sea with period $J$ around the circle. One can expect that the continuum description in terms of collective behavior of the eigenvalues will start 
to break down when we can resolve the individual eigenvalues near the top of the Fermi 
sea. If we choose some cutoff thickness near the top of the sea, which is given by the 
thickness of the wave functions, (namely of order $1$), then the number of eigenvalues at 
the top is of order $\sqrt N$. This means that for string states of angular momentum of 
order $J\sim \sqrt N$ the perturbation theory in $1/N^2$ should start to converge badly, because we can not ignore the granularity of the eigenvalues beyond this point.
This is exactly what is observed in the direct calculations, and resonates with the 
observations of Shenker on the nature of non-perturbative effects in matrix models being always related to the dynamics of a single eigenvalue \cite{S}. Indeed, adding one eigenvalue costs $N$ in energy for the ground state. This is of order $1/g_s\sim 1/(1/N)$, as expected 
for the tension of a D-brane.

We can also ask if this matrix quantum mechanical 
model arises naturally in some other  string theory than ${\cal N}=4$ SYM theory. 
We understand this
 intuition for the $c=1$ matrix model \cite{McV}, so it would be nice to have a similar description of the setup presented in this paper. One could conjecture that 
 there are solutions of string theory where one can place D0-branes in a potential well where there is one almost massless modulus for the D-branes, and there is no need to require the background where the D-branes are located to be supersymmetric. 
 The low energy limit of the open string theory on these D-branes would look as
 a gauged matrix quantum mechanics with some potential along this flat direction, which would begin with a quadratic term. Then we could hope that the near horizon geometry of these 
 D0-branes would be holographically dual to the gauged matrix model above. 
 
 For example, Verlinde \cite{VS} has shown that in a supersymmetric gauged conformal quantum mechanics there is a time slicing of $AdS_2$ which leads to a quadratic 
 potential for eigenvalues plus potential terms that relate different eigenvalues and depend on some fermion occupation numbers. 
  These extra potential terms in the eigenvalues that can be turned 
 off by requiring the fermionic components of the supereigenvalues to be in their ground states
 (set $\kappa_{ij}= 1 $ in equation 55 of his paper).
 This sector coincides with the matrix model studied in this paper, and it seems to arise from a completely different physical system which does live in $1+1$ dimensions.

\section{Conclusion}

We have argued in this paper that the gauged Gaussian matrix quantum mechanics has the potential to describe
a string theory which is exactly solvable, much in the spirit of the $c=1$ matrix model, and which behaves very much like the AdS/CFt correspondence.

We have presented strong evidence for this claim: 
we have shown that in this model we can  explicitly describe the 
open-closed string duality, by relating the eigenvalues of the matrix model to 
the closed string states made out of traces of the fields. This depends on a gauge choice
of how we choose to solve the model.
Moreover, we have also shown that the system arises as a decoupling limit of ${\cal N}=4$ SYM theory, by taking a different time slicing of the associated AdS spacetime.
From this point of view we have been able to show that
the eigenvalues of the matrix model
behave as D-branes, as well as holes in the Fermi sea of eigenvalues. 
This has been motivated 
by a comparison between this model and the operators which describe half-BPS 
giant gravitons in $AdS_5\times S^5$. Also, this comparison shows that the giant 
gravitons have an interpretation in terms of the dynamics of a single eigenvalue of 
the dual SYM theory. This leaves us with a possibility to study the giant 
semiclassically both in the supergravity and the SYM theory. Previously, the semiclassical description in SYM  was missing (although this description was hinted at in \cite{HHI}), so at least from this point of view we have learned a valuable piece of information.
We have also given some new interpretations to the structure of the non-planar 
$J^2/N$ expansion in the plane wave limit. We have argued that this dependence on 
$J$ is natural if we think of it as arising
from the breakdown of the collective behavior of the eigenvalues, due to
the granularity of the edge of the Fermi 
sea of the matrix model.

We do not claim to have a geometric dual description because the dual  geometry seems to be too strongly curved to provide a semiclassical picture. This in itself should not worry us too much because we have found other examples in string theory, like Gepner models, where a clear 
interpretation of the geometry is missing. In those cases we are still content with the explicit 
solution of the string spectrum, and are perfectly happy to call the space time a stringy geometry. Similarly, here we can be content with the same type of description. 
Based on minimal assumptions, we have  guessed 
that the dual geometry is two dimensional with an  AdS like boundary, 
but beyond that we do not have more information on the geometry.

It would be very interesting to understand how this matrix model is related to the $c=1$ 
matrix model in more detail. Naively, the change in the potential from $- X^2/\alpha'$ to
$X^2/\alpha'$ looks like an analytic continuation where $\alpha'\to -\alpha'$. However, here 
the only limit we have is t' Hooft's large $N$ limit and there is no double scaling limit. The number of eigenvalues on both theories is very different, in the $c=1$ theory we take $N\to\infty$ strictly, here $N$ is finite but very large. Indeed, 
$N$ being fininte was very important to the description of certain aspects of the 
D-brane physics in terms of Young tableaux.

Probably the most important open problem for this model is to find a 
target space geometry for the string theory that describes it. Seeing as it arises from a decoupling limit of string theory in $AdS_5\times S^5$, one suspects that there should be a geometrical description that captures this limit and no additional states.

\section*{Acknowledgements}

It is a pleasure to thank V. Balasubramanian, B. Feng,
D. Gross, M. Huang,  I. Klebanov, O. Lunin, J. Maldacena, J. Polchisnki, N. Seiberg, 
H. Verlinde, E. Witten for various discussions related to this work. 
Work supported in part by DOE Grant No.~DE-FG02-90ER40542.


\begin{thebibliography}{99}







%\cite{McGreevy:2003kb}
\bibitem{McV}
J.~McGreevy and H.~Verlinde,
``Strings from tachyons: The c = 1 matrix reloaded,''
JHEP {\bf 0312}, 054 (2003)
[arXiv:hep-th/0304224].
%%CITATION = HEP-TH 0304224;%%




%\cite{McGreevy:2003ep}
\bibitem{McTV}
J.~McGreevy, J.~Teschner and H.~Verlinde,
``Classical and quantum D-branes in 2D string theory,''
JHEP {\bf 0401}, 039 (2004)
[arXiv:hep-th/0305194].
%%CITATION = HEP-TH 0305194;%

\bibitem{Sen}
A.~Sen,
``Rolling tachyon,''
JHEP {\bf 0204}, 048 (2002)
[arXiv:hep-th/0203211].
%%CITATION = HEP-TH 0203211;%%


%\cite{Fateev:2000ik}
\bibitem{FZZ}
V.~Fateev, A.~B.~Zamolodchikov and A.~B.~Zamolodchikov,
 ``Boundary Liouville field theory. I: Boundary state and boundary  two-point
function,''
arXiv:hep-th/0001012.
%%CITATION = HEP-TH 0001012;%%


%\cite{Klebanov:2003km}
\bibitem{KMS}
I.~R.~Klebanov, J.~Maldacena and N.~Seiberg,
``D-brane decay in two-dimensional string theory,''
JHEP {\bf 0307}, 045 (2003)
[arXiv:hep-th/0305159].
%%CITATION = HEP-TH 0305159;%%

%\cite{Takayanagi:2003sm}
\bibitem{TT}
T.~Takayanagi and N.~Toumbas,
``A matrix model dual of type 0B string theory in two dimensions,''
JHEP {\bf 0307}, 064 (2003)
[arXiv:hep-th/0307083].
%%CITATION = HEP-TH 0307083;%%




%\cite{Douglas:2003up}
\bibitem{DKKMMS}
M.~R.~Douglas, I.~R.~Klebanov, D.~Kutasov, J.~Maldacena, E.~Martinec and N.~Seiberg,
``A new hat for the c = 1 matrix model,''
arXiv:hep-th/0307195.
%%CITATION = HEP-TH 0307195;%%

%\cite{Polchinski:1994jp}
\bibitem{P}
J.~Polchinski,
``On the nonperturbative consistency of d = 2 string theory,''
Phys.\ Rev.\ Lett.\  {\bf 74}, 638 (1995)
[arXiv:hep-th/9409168].
%%CITATION = HEP-TH 9409168;%%

%\cite{Klebanov:1991qa}
\bibitem{Kleb}
I.~R.~Klebanov,
``String theory in two-dimensions,''
arXiv:hep-th/9108019.
%%CITATION = HEP-TH 9108019;%%

%\cite{Mukhi:2003sz}
\bibitem{Muk}
S.~Mukhi,
``Topological matrix models, Liouville matrix model and c = 1 string theory,''
arXiv:hep-th/0310287.
%%CITATION = HEP-TH 0310287;%%


%\cite{Brezin:1977sv}
\bibitem{BIPZ}
E.~Brezin, C.~Itzykson, G.~Parisi and J.~B.~Zuber,
``Planar Diagrams,''
Commun.\ Math.\ Phys.\  {\bf 59}, 35 (1978).
%%CITATION = CMPHA,59,35;%%


%\cite{'tHooft:1973jz}
\bibitem{tH}
G.~'t Hooft,
``A Planar Diagram Theory For Strong Interactions,''
Nucl.\ Phys.\ B {\bf 72}, 461 (1974).
%%CITATION = NUPHA,B72,461;%%


%\cite{Maldacena:1997re}
\bibitem{M}
J.~M.~Maldacena,
``The large N limit of superconformal field theories and supergravity,''
Adv.\ Theor.\ Math.\ Phys.\  {\bf 2}, 231 (1998)
[Int.\ J.\ Theor.\ Phys.\  {\bf 38}, 1113 (1999)]
[arXiv:hep-th/9711200].
%%CITATION = HEP-TH 9711200;%%

%\cite{Witten:1998qj}
\bibitem{W}
E.~Witten,
``Anti-de Sitter space and holography,''
Adv.\ Theor.\ Math.\ Phys.\  {\bf 2}, 253 (1998)
[arXiv:hep-th/9802150].
%%CITATION = HEP-TH 9802150;%%

%\cite{Gubser:1998bc}
\bibitem{GKP}
S.~S.~Gubser, I.~R.~Klebanov and A.~M.~Polyakov,
``Gauge theory correlators from non-critical string theory,''
Phys.\ Lett.\ B {\bf 428}, 105 (1998)
[arXiv:hep-th/9802109].
%%CITATION = HEP-TH 9802109;%%



%\cite{Gopakumar:2003ns}
\bibitem{G1}
R.~Gopakumar,
``From free fields to AdS,''
arXiv:hep-th/0308184.
%%CITATION = HEP-TH 0308184;%%

%\cite{Gopakumar:2004qb}
\bibitem{G2}
R.~Gopakumar,
``From free fields to AdS. II,''
arXiv:hep-th/0402063.
%%CITATION = HEP-TH 0402063;%%

%\cite{Strominger:2003tm}
\bibitem{St}
A.~Strominger,
``A matrix model for AdS(2),''
arXiv:hep-th/0312194.
%%CITATION = HEP-TH 0312194;%%

%\cite{Verlinde:2004gt}
\bibitem{VS}
H.~Verlinde,
``Superstrings on AdS(2) and superconformal matrix quantum mechanics,''
arXiv:hep-th/0403024.
%%CITATION = HEP-TH 0403024;%%

%\cite{Ho:2004qp}
\bibitem{Ho}
P.~M.~Ho,
``Isometry of AdS(2) and the c = 1 matrix model,''
arXiv:hep-th/0401167.
%%CITATION = HEP-TH 0401167;%%

%\cite{Khoury:2000hz}
\bibitem{KV}
J.~Khoury and H.~Verlinde,
``On open/closed string duality,''
Adv.\ Theor.\ Math.\ Phys.\  {\bf 3}, 1893 (1999)
[arXiv:hep-th/0001056].
%%CITATION = HEP-TH 0001056;%%




%\cite{Kristjansen:2002bb}
\bibitem{KPSS}
C.~Kristjansen, J.~Plefka, G.~W.~Semenoff and M.~Staudacher,
 ``A new double-scaling limit of N = 4 super Yang-Mills theory and PP-wave
strings,''
Nucl.\ Phys.\ B {\bf 643}, 3 (2002)
[arXiv:hep-th/0205033].
%%CITATION = HEP-TH 0205033;%%


%\cite{Berenstein:2002jq}
\bibitem{BMN}
D.~Berenstein, J.~M.~Maldacena and H.~Nastase,
``Strings in flat space and pp waves from N = 4 super Yang Mills,''
JHEP {\bf 0204}, 013 (2002)
[arXiv:hep-th/0202021].
%%CITATION = HEP-TH 0202021;%%



%\cite{Berenstein:2002sa}
\bibitem{BN}
D.~Berenstein and H.~Nastase,
``On lightcone string field theory from super Yang-Mills and holography,''
arXiv:hep-th/0205048.
%%CITATION = HEP-TH 0205048;%%



%\cite{Constable:2002hw}
\bibitem{C7}
N.~R.~Constable, D.~Z.~Freedman, M.~Headrick, S.~Minwalla, L.~Motl, A.~Postnikov and W.~Skiba,
``PP-wave string interactions from perturbative Yang-Mills theory,''
JHEP {\bf 0207}, 017 (2002)
[arXiv:hep-th/0205089].
%%CITATION = HEP-TH 0205089;%%


%\cite{Corley:2001zk}
\bibitem{CJR}
S.~Corley, A.~Jevicki and S.~Ramgoolam,
``Exact correlators of giant gravitons from dual N = 4 SYM theory,''
Adv.\ Theor.\ Math.\ Phys.\  {\bf 5}, 809 (2002)
[arXiv:hep-th/0111222].
%%CITATION = HEP-TH 0111222;%%




%\cite{Gross:1992tu}
\bibitem{G}
D.~J.~Gross,
``Two-dimensional QCD as a string theory,''
Nucl.\ Phys.\ B {\bf 400}, 161 (1993)
[arXiv:hep-th/9212149].
%%CITATION = HEP-TH 9212149;%%

%\cite{Minahan:1993np}
\bibitem{MP}
J.~A.~Minahan and A.~P.~Polychronakos,
``Equivalence of two-dimensional QCD and the C = 1 matrix model,''
Phys.\ Lett.\ B {\bf 312}, 155 (1993)
[arXiv:hep-th/9303153].
%%CITATION = HEP-TH 9303153;%%


%\cite{Jevicki:1996wu}
\bibitem{JvT}
A.~Jevicki and A.~van Tonder,
``Finite [q-Oscillator] Description of 2-D String Theory,''
Mod.\ Phys.\ Lett.\ A {\bf 11}, 1397 (1996)
[arXiv:hep-th/9601058].
%%CITATION = HEP-TH 9601058;%%

%\cite{Polychronakos:1996rj}
\bibitem{Poly}
A.~P.~Polychronakos,
``Quasihole wavefunctions for the Calogero model,''
Mod.\ Phys.\ Lett.\ A {\bf 11}, 1273 (1996)
[arXiv:cond-mat/9603132].
%%CITATION = COND-MAT 9603132;%%


%\cite{McGreevy:2000cw}
\bibitem{McST}
J.~McGreevy, L.~Susskind and N.~Toumbas,
``Invasion of the giant gravitons from anti-de Sitter space,''
JHEP {\bf 0006}, 008 (2000)
[arXiv:hep-th/0003075].
%%CITATION = HEP-TH 0003075;%%




%\cite{Grisaru:2000zn}
\bibitem{GMT}
M.~T.~Grisaru, R.~C.~Myers and O.~Tafjord,
``SUSY and Goliath,''
JHEP {\bf 0008}, 040 (2000)
[arXiv:hep-th/0008015].
%%CITATION = HEP-TH 0008015;%%

%\cite{Hashimoto:2000zp}
\bibitem{HHI}
A.~Hashimoto, S.~Hirano and N.~Itzhaki,
``Large branes in AdS and their field theory dual,''
JHEP {\bf 0008}, 051 (2000)
[arXiv:hep-th/0008016].
%%CITATION = HEP-TH 0008016;%%


%\cite{Balasubramanian:2001nh}
\bibitem{BBNS}
V.~Balasubramanian, M.~Berkooz, A.~Naqvi and M.~J.~Strassler,
``Giant gravitons in conformal field theory,''
JHEP {\bf 0204}, 034 (2002)
[arXiv:hep-th/0107119].
%%CITATION = HEP-TH 0107119;%%




%\cite{Banks:1996vh}
\bibitem{BFSS}
T.~Banks, W.~Fischler, S.~H.~Shenker and L.~Susskind,
``M theory as a matrix model: A conjecture,''
Phys.\ Rev.\ D {\bf 55}, 5112 (1997)
[arXiv:hep-th/9610043].
%%CITATION = HEP-TH 9610043;%%

%\cite{Gaiotto:2003yf}
\bibitem{GIR}
D.~Gaiotto, N.~Itzhaki and L.~Rastelli,
``On the BCFT description of holes in the c = 1 matrix model,''
Phys.\ Lett.\ B {\bf 575}, 111 (2003)
[arXiv:hep-th/0307221].
%%CITATION = HEP-TH 0307221;%%

%\cite{Aharony:2002nd}
\bibitem{AABF}
O.~Aharony, Y.~E.~Antebi, M.~Berkooz and R.~Fishman,
 ``'Holey sheets': Pfaffians and subdeterminants as D-brane operators in large
N gauge theories,''
JHEP {\bf 0212}, 069 (2002)
[arXiv:hep-th/0211152].
%%CITATION = HEP-TH 0211152;%%





%\cite{Berenstein:2003ah}
\bibitem{B}
D.~Berenstein,
``Shape and holography: Studies of dual operators to giant gravitons,''
Nucl.\ Phys.\ B {\bf 675}, 179 (2003)
[arXiv:hep-th/0306090].
%%CITATION = HEP-TH 0306090;%%


%\cite{Gubser:2002tv}
\bibitem{GKP2}
S.~S.~Gubser, I.~R.~Klebanov and A.~M.~Polyakov,
``A semi-classical limit of the gauge/string correspondence,''
Nucl.\ Phys.\ B {\bf 636}, 99 (2002)
[arXiv:hep-th/0204051].
%%CITATION = HEP-TH 0204051;%%




%\cite{Balasubramanian:2002sa}
\bibitem{BHLN}
V.~Balasubramanian, M.~x.~Huang, T.~S.~Levi and A.~Naqvi,
``Open strings from N = 4 super Yang-Mills,''
JHEP {\bf 0208}, 037 (2002)
[arXiv:hep-th/0204196].
%%CITATION = HEP-TH 0204196;%%



%\cite{Shenker:1990uf}
\bibitem{S}
S.~H.~Shenker,
``The Strength Of Nonperturbative Effects In String Theory,''
RU-90-47
%\href{http://www.slac.stanford.edu/spires/find/hep/www?r=ru-90-47}{SPIRES entry}
{\it Presented at the Cargese Workshop on Random Surfaces, Quantum Gravity and Strings, Cargese, France, May 28 - Jun 1, 1990}



%\cite{BKR}
\bibitem{BKR}
A.~Boyarsky, B.~Kulik and O.~Ruchayskiy,
 ``Classical and quantum branes in c = 1 string theory and quantum Hall effect,''
arXiv:hep-th/0312242.
%%CITATION = HEP-TH 0312242;%%


\end{thebibliography}
 \end{document}